\documentclass[aip,reprint]{revtex4-1}

\usepackage{graphicx}
\usepackage{dcolumn}
\usepackage{bm}

\usepackage[utf8]{inputenc}
\usepackage[T1]{fontenc}
\usepackage{mathptmx}
\usepackage{etoolbox}
\usepackage{color}
\usepackage{amsmath}

\makeatletter
\def\@email#1#2{%
 \endgroup
 \patchcmd{\titleblock@produce}
  {\frontmatter@RRAPformat}
  {\frontmatter@RRAPformat{\produce@RRAP{*#1\href{mailto:#2}{#2}}}\frontmatter@RRAPformat}
  {}{}
}%
\makeatother
\begin{document}

\preprint{AIP/123-QED}

\title[]{Stochastic resonance in higher-order networks driven by colored noise}
\author{Zhongwen Bi}
\affiliation{School of Mathematics and Statistics, Northwestern Polytechnical University, Xi'an 710072, China}
\affiliation{MOE Key Laboratory for Complexity Science in Aerospace, Northwestern Polytechnical University, Xi’an 710072, China}
\author{Dan Zhao}%
\affiliation{ 
Potsdam Institute for Climate Impact Research, Potsdam 14412, Germany}%
\affiliation{Department of Physics, Humboldt University Berlin, Berlin 12489, Germany}

\author{Qi Liu}
\affiliation{ 
Department of Systems and Control Engineering, Institute of Science Tokyo (former Tokyo Tech), Tokyo 152-8552, Japan}

\author{J\"urgen Kurths}
\affiliation{ 
Potsdam Institute for Climate Impact Research, Potsdam 14412, Germany}%
\affiliation{Department of Physics, Humboldt University Berlin, Berlin 12489, Germany}

\author{Yong Xu}
\email{hsux3@nwpu.edu.cn.}
\affiliation{School of Mathematics and Statistics, Northwestern Polytechnical University, Xi'an 710072, China}
\affiliation{MOE Key Laboratory for Complexity Science in Aerospace, Northwestern Polytechnical University, Xi’an 710072, China}%

\date{\today}

\begin{abstract}
We investigate stochastic resonance (SR) in an ensemble of coupled overdamped bistable oscillators driven by colored noise. The networks incorporate the weighted contributions of both pairwise coupling and 2-simplex coupling. Our findings show that these higher-order interactions further exacerbate the suppression effect of colored noise on SR, reducing the peak of resonance curves and shifting the optimal noise intensity toward higher values. To clarify the underlying mechanism, we establish a close connection between SR and the four-stage variation in network synchronization level. Specially, the synchronization extremes explain the effect of higher-order coupling and colored noise on SR. Our analysis reveals that higher-order interactions do not reverse, but primarily promote the spatial propagation of suppression effects due to colored noise.
\end{abstract}

\maketitle

\begin{quotation}
Suppression effects of stochastic resonance (SR) by colored noise is widely accepted. Such effects have been observed both in one-dimensional overdamped bistable oscillators and in pairwise-coupled oscillators. However, this suppression depends on the characteristics of the nonlinear systems. Recently, numerous studies have demonstrated that the introduction of higher-order interactions substantially alters network dynamics. This raises the question: Could higher-order coupling also reverse suppression effects of colored noise?
In this work, we find that the presence of higher-order interactions does not reverse, but instead exacerbates, suppression effects of colored noise. The basic higher-order coupling scheme considered in this study facilitates the spatial propagation of the suppression effect, 
leading to a conjecture: it is possible to discover an example of color noise enhancing SR in the collective dynamics of overdamped bistable oscillators by designing an ingenious higher-order coupling strategy.
\end{quotation}

\section{Introduction}

Stochastic resonance (SR) is a basic noise-induced resonance phenomenon in nonlinear systems, whereby a weak periodic signal can be optimally amplified by an appropriate level of random fluctuations \cite{benzi1981mechanism}. SR has been extensively studied in a wide range of physical \cite{zhou2003noise, xu2012stochastic,xu2013levy}, biological \cite{wang2016levy}, and engineered settings \cite{li2024enhanced,liu2022complex,liu2023complex}, and its theoretical foundations and applications have been comprehensively reviewed \cite{gammaitoni1998stochastic}.

Although SR phenomena in one-dimensional nonlinear systems have been well understood, coupled systems give rise to a number of attractive collective behaviors. These include array-enhanced SR \cite{lindner1995array}, system size  resonance \cite{pikovsky2002system}, and diversity-induced  resonance \cite{tessone2006diversity}, and a substantial body of related works \cite{jung1992collective, pikovsky1997coherence, zhou2002spatiotemporal,liu2025flutter}. These network-specific collective SR phenomena rely on the rich high-dimensional parameter space of coupled networks.

However, the existing studies on SR in networks are limited to pairwise coupling, without including of higher-order coupling. Pairwise coupling encodes two-body interactions, whereas higher-order coupling characterizes three-body or many-body interactions. In reality, many networks exhibit higher-order interactions, including social networks \cite{cencetti2021temporal}, ecological networks \cite{grilli2017higher}, and brain functional networks \cite{petri2014homological}. Recent extensive studies have revealed that higher-order interactions shape collective dynamics \cite{battiston2020networks,zhang2023higher,millan2025topology, muolo2023turing}, including but not limited to chaos \cite{bick2016chaos}, synchronization \cite{gallo2022synchronization,parastesh2022synchronization}, desynchronization \cite{skardal2019abrupt}, synchronization transitions\cite{zhao2025synchronization}, and explosive synchronization \cite{malizia2025hyperedge}. Furthermore,  higher-order interactions introduce nonlinearity at the macroscopic level \cite{skardal2020higher}, and rich higher-order coupling representations also influence synchronization dynamics \cite{boccaletti2023structure,bick2023higher,benson2016higher}. Despite this rapid progress in higher-order network science, SR in higher-order networks has received comparatively less attention. Existing studies have reported SR in higher-order coupled bistable oscillators \cite{semenov2025nonlocal,wang2026higher} and coupled phase oscillators \cite{wang2025network} under Gaussian white noise, and have explored SR in small-world networks incorporating higher-order neural-motif interactions \cite{li2024stochastic}.

 Although white noise possesses very convenient mathematical properties for analysis, it does not always appropriately describe many stochastic perturbations that are temporally correlated. Non-Markovian colored noise has been observed in physics \cite{haunggi1994colored, gomes2020mean,xu2011stochastic}, neuroscience \cite{peng1993long,teich2002fractal}, quantum mechanics \cite{virzi2022quantum,long2022entanglement}, and engineered settings \cite{zhang2025tipping,zhao2023occurrence,liu2018sliding,zhang2021rate}. The one-dimensional bistable overdamped system driven by additive colored noise has been theoretically analyzed \cite{hanggi1993can}: increasing the correlation time suppresses the peak of SR and requires a larger noise intensity. The effect of colored noise on SR depends on the properties of the nonlinear system. Neuronal systems exhibit an enhancement at local \cite{soma20031} or global noise intensities \cite{nozaki1998enhancement}. Mixed colored noise \cite{jia2001effects} and non-Gaussain colored noise \cite{fuentes2001enhancement} can also enhance SR in bistable systems. In addition, the finite damping in the bistable system regulates the suppression effect \cite{gammaitoni1989periodically}. A recent study \cite{liu2025optimizing} has reported the emergence of suppression as well as enhancement effects in colored noise driven overdamped and underdamped bistable systems with pairwise coupling form, respectively. 

These observations motivate the central question of this work: Does higher-order coupling modify the canonical suppression effect of colored noise on SR in coupled overdamped bistable networks? To answer this question, we investigate SR in higher-order networks with triadic interactions driven by temporally correlated Gaussian noise, and we explicitly compare two standard Ornstein--Uhlenbeck parameterizations (intensity-normalized versus power-limited), as defined in the Section \ref{sec:sec2}. In Section \ref{sec:sec3}, we characterize how the SR profile, including the peak amplification and the optimal noise level, depends on the noise correlation time, higher-order interaction weight, and coupling strength. Additionally, in Section \ref{sec:sec4}, we connect these macroscopic trends to microscopic dynamics by analyzing the mean-field switching frequency and the spatiotemporal synchronization level of the networks.

\section{Model description}
\label{sec:sec2}
We consider an ensemble of coupled overdamped bistable oscillators, governed by the following stochastic differential equations,
\begin{equation}\label{eq:eq1}
\frac{\mathrm{d}x_i}{\mathrm{d}t}=f(x_i)+A\sin(\Omega t)+G(\boldsymbol{x},x_i)+\zeta_i(t), 
\end{equation}
in which $\boldsymbol{x}=(x_1,x_2,...,x_N)$, $x_i$ denotes the response of the $i$-th oscillator, and its evolution is determined by four contributions: the intrinsic dynamics $f(x_i)=ax_i-bx_i^3$, the interactions $G(\boldsymbol{x},x_i)$, the external periodic forcing $A\sin(\Omega t)$, and the random noise $\zeta_i(t)$. Without loss of generality, we set $a = b = 1$. The intrinsic dynamics is then $f(x_i)=x_i-x_i^3$, which corresponds to a symmetric double-well potential $P(x)= -1/2x^{2} + 1/4x^{4}$, with a potential barrier between two symmetric potential wells. The bottom of the potential well is located at $x_p=\pm 1$ and the height of the potential barrier is $1/4$. There is an unstable point ($x=0$) between the two potential wells.

The interactions between nodes are a convex combination of triadic and pairwise couplings, weighted by $\alpha$ and $1-\alpha$,
\begin{equation}
\begin{aligned}
G(\boldsymbol{x},x_i)
&= \alpha \frac{\sigma}{K}\sum_{j,k}^{N}\frac{1}{2}B_{ijk}\,
\tanh\!\big(x_j+x_k-2x_i\big) \\
&\quad + (1-\alpha)\frac{\sigma}{N}\sum_{j}^{N}\big(x_j-x_i\big).
\end{aligned}
\end{equation}
in which $\sigma$ is the overall coupling strength and $ N$ is the system size. The hyperbolic tangent function $\tanh(x_j+x_k -2x_i)$ captures the group effect of three nodes and is symmetric under permutation of $j$ and $k$. $B_{ijk}=1$ if $i$, $j$, and $k$ are mutually connected (i.e., form a simplex), and $B_{ijk}=0$ otherwise. The higher-order degree of the $i$-th oscillator is defined by
$K_i=\frac{1}{2}\sum_{j,k=1}^{N}B_{ijk}$, and for globally coupled networks considered in this study, all nodes are equivalent, so $K=K_i=(N-1)(N-2)/2$ for $i=1,2,...,N$. The parameter $\alpha$ tunes the relative weight of the two interaction channels: at $\alpha = 0$, the system consists exclusively of pairwise interactions, whereas at $\alpha = 1$, it becomes purely triadic couplings. In the following, we vary $\alpha\in[0,1]$ to quantify how the higher-order interactions influence the collective responses.

We consider two colored-noise models based on the Ornstein--Uhlenbeck (OU) process. They differ in the normalization of the noise amplitude with respect to the correlation time $\tau$, such that the the parameter $D$ has distinct physical interpretations in both cases.

\noindent\textbf{Type-I: Intensity-normalized OU colored noise (fixed white-noise limit) \cite{jung1988bistability}.}
The colored noise $\zeta_i(t)$ is generated by
\begin{equation}
\mathrm{d}\zeta_i(t) = -\frac{1}{\tau}\zeta_i(t)\,\mathrm{d}t + \sqrt{\frac{2D}{\tau^{2}}}\,\mathrm{d}W_i(t),
\end{equation}
and
\begin{equation}
E( \zeta_i(t)\zeta_j(t'))=\frac{D}{\tau}\exp\!\left(-\frac{|t-t'|}{\tau}\right)\delta_{ij},
\end{equation}
where $W_i(t)$ is standard Brownian motion, $E(\cdot)$ is mathematical expectation operator. $E( \zeta_i(t)\zeta_j(t')) \to 2D\delta_{ij}\delta(t-t')$ as $\tau\to0$, 
so $D$ directly fixes the intensity of the effective white-noise limit.

\noindent\textbf{Type-II: Power-limited OU colored noise (fixed variance).}
In the power-limited case, the OU process is normalized to keep the stationary variance equal to $D^{\mathrm{pl}}$ for all $\tau$ \cite{ma2008coherence, jung2005thermal}:
\begin{equation}\label{eq:eq3}
\mathrm{d}\zeta_i^{\mathrm{pl}}(t)=-\frac{1}{\tau}\zeta_i^{\mathrm{pl}}(t)\,\mathrm{d}t+\sqrt{\frac{2D^{\mathrm{pl}}}{\tau}}\,\mathrm{d}W_i(t),
\end{equation}
and
\begin{equation}
E( \zeta_i^{\mathrm{pl}}(t)\zeta_j^{\mathrm{pl}}(t'))=
D^{\mathrm{pl}}\exp\!\left(-\frac{|t-t'|}{\tau}\right)\delta_{ij},
\end{equation}
such that $E(\zeta_i^{\mathrm{pl}})^2)=D^{\mathrm{pl}}$ is independent of $\tau$. 
Accordingly, $D^{\mathrm{pl}}$ should be interpreted as the noise variance (instantaneous power), rather than the effective white-noise intensity in the $\tau\to0$ limit.

Both models represent the same OU family but with different parameterizations. 
When comparing the two autocorrelation functions at the same $\tau$, they coincide if \(D^{\mathrm{pl}}=D/\tau\). 
We refer to Type-I as classical colored noise. SR is closely related to a noise-induced escape from the potential wells, and these two parameterizations can have qualitatively different effects on the escape behavior. 
The correlation time of classical colored noise monotonically suppresses the escape rate \cite{jung1988bistability}, whereas power-limited colored noise can facilitate or suppress it in a non-monotonic manner \cite{jung2005thermal}. 
In this work, we focus primarily on SR driven by classical colored noise; power-limited colored noise is used here only for comparison in selected cases. 
Unless otherwise specified, the noise type is classical colored noise. 

For the simulations, we set the time step $\Delta t=0.01$ and the total integration time $T=2000$. 
The initial conditions $x_{i}(0)$ are independently drawn from the uniform distribution on [-1,1], our results are averaged over 50 independent realizations. 
The mean-field variable is $X(t)=\frac{1}{N}\sum_{i=1}^{N}x_{i}(t)$, whose SR is quantified by the spectral amplification factor \cite{jung1991amplification,pikovsky2002system}, $S=4({|M_1|}/{A}) ^2$, where $M_1$ is the power spectral density evaluated at  the frequency $\Omega$ of the external periodic force.

\section{SR: Profiles and parameter dependence}\label{sec:sec3}
\begin{figure}[!htb]
	\centering
	\includegraphics[width=0.45\columnwidth]{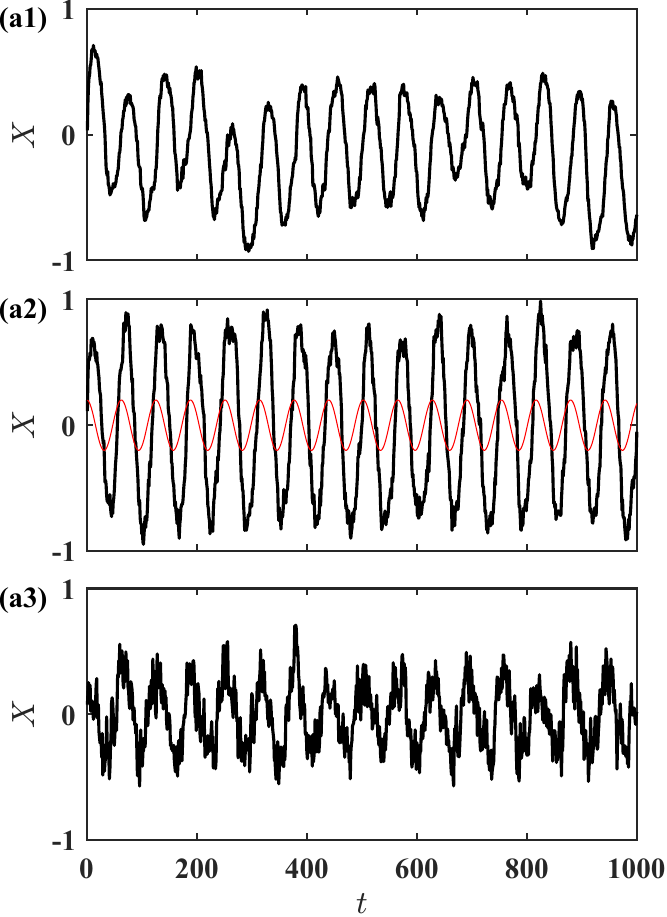}\quad
    \includegraphics[width=0.45\columnwidth]{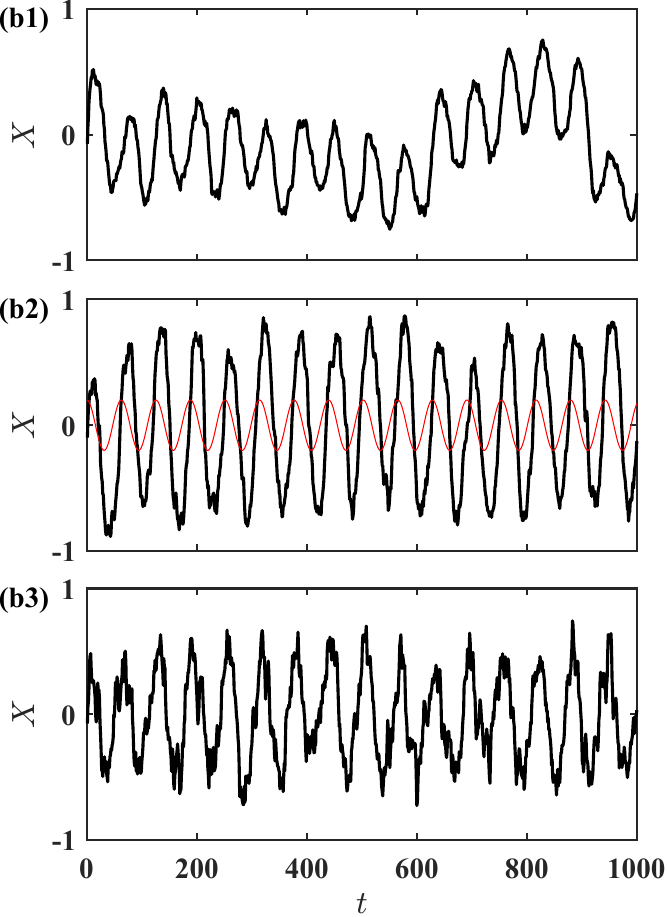}
	\caption{Mean-field time series $X(t)$ on the higher-order network (1) for different noise intensities $D$. Parameters: $A=0.2$, $\Omega=0.1$, $N=50$, $\alpha=1$, $\sigma=0.1$. From top to bottom, $D=0.06,\,0.16,\,0.60$. (a) White noise (white-noise limit $\tau=0$); (b) Colored noise ($\tau=0.5$). The red curves in panels (a2) and (b2) show the external periodic forcing $A\sin(\Omega t)$.}
    \label{fig:fig1}
\end{figure}

We first characterize the collective responses of the higher-order networks to the external periodic excitation by monitoring the mean-field variable $X(t)$. 
From the time-domain perspective, Fig.~\ref{fig:fig1} shows representative realizations of $X(t)$ for different noise intensities $D$, comparing the white-noise limit ($\tau=0$) with colored noise ($\tau>0$).

For $\tau=0$, Figs.~\ref{fig:fig1}(a1)--(a3) show three representative regimes. 
For weak noise $D=0.06$, the system switches randomly between the two potential wells at $x_p=\pm1$. 
Note that $X(t)$ does not always stay closely at $x_p=-1$. 
Such behavior arises from the averaging effect inherent in the mean-field response. 
In particular, barrier crossing is a microscopic event: at a given time only a subset of oscillators may overcome the barrier, while the rest remain trapped. 
Since the fraction of oscillators that cross the barrier fluctuates in time and is generally less than 1, $X(t)$ takes intermediate values in $[-1,1]$. 

When $D=0.16$, $X(t)$ displays a clear periodicity and becomes phase synchronized with the external periodic force, indicating an enhanced phase-locked collective response driven by the synergy between noise-controlled switching and higher-order coupling. 
For a larger $D$, e.g. $D=0.6$, noise becomes the dominant factor. 
$X(t)$ then shows pronounced randomness and is reduced.

Figures \ref{fig:fig1}(b1)--(b3) present the clearly colored noise case of $\tau=0.5$ for different $D$. 
The behaviors across different $D$ are qualitatively similar to that in Figs.~\ref{fig:fig1}(a1)--(a3): as $D$ increases, $X(t)$ transitions from a phase-locked (nearly) periodic response to a noise-dominated random state. 
However, a clear quantitative difference emerges at weak noise: for the same small noise intensity ($D=0.06$), the colored noise induces markedly fewer barrier-crossing events than white noise. 
This suggests that temporal correlations effectively smooth the noise over short time scales, reducing the probability of sufficiently large instantaneous excursions that trigger switching, and thereby requiring a larger $D$ to achieve comparable inter-well transitions.

\begin{figure}[!htb]
    \centering
	\includegraphics[width=0.45\columnwidth]{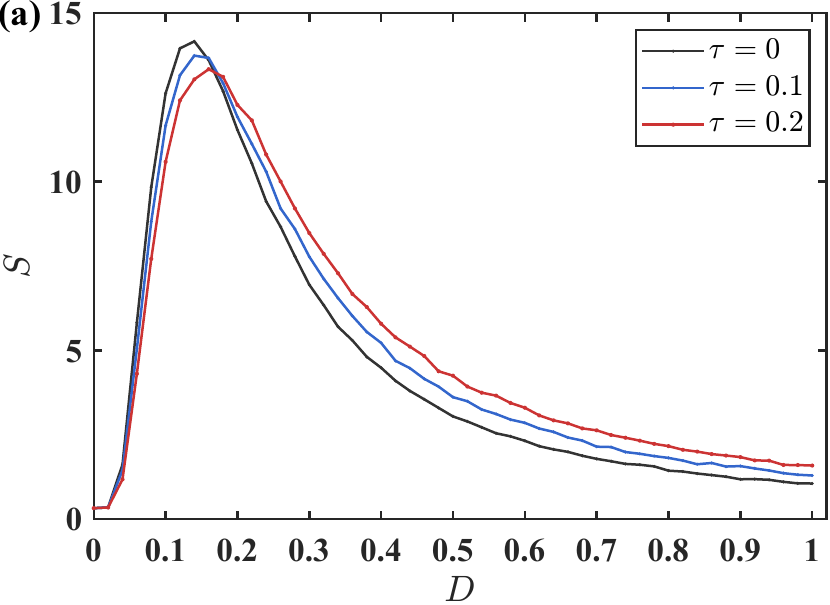}\quad
    \includegraphics[width=0.485\columnwidth]{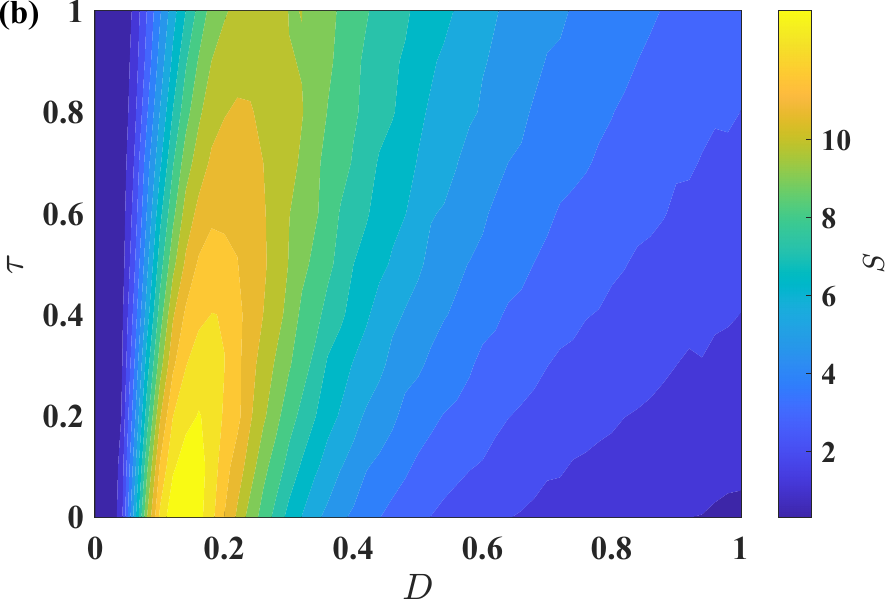}
	\caption{Spectral amplification factor $S$ versus noise intensity $D$ for different correlation time $\tau$ on the higher-order network (1), $\alpha=1$. (a) Representative curves for $\tau=0,\,0.1,\,0.2$. (b) Heat map of $S$ in the parameter plane $(D,\tau)$. Other parameters: $A=0.2$, $\Omega=0.1$, $N=50$, $\sigma=0.1$.}
    \label{fig:fig2}
\end{figure}

Following the above analysis, we compute the spectral amplification factor $S$ for Eq.~(\ref{eq:eq1}), with the results presented in Fig.~\ref{fig:fig2}. 
Figure~\ref{fig:fig2}(a) shows that as $D$ increases, $S$ exhibits a non-monotonic dependence on $D$: it first increases and then decreases. 
This indicates the existence of an optimal noise intensity $D_{\mathrm{optimal}}$ at which $S$ attains its maximum, i.e. the resonance peak $S_{\mathrm{peak}}$. 
When $\tau=0$, $S_{\mathrm{peak}}$ is the largest and the corresponding $D_{\mathrm{optimal}}$ is the smallest. 
As $\tau$ increases, the resonance peak is reduced, while $D_{\mathrm{optimal}}$ shifts to larger values. 
These trends quantitatively confirm the time-series observations in Fig.~\ref{fig:fig1}: stronger temporal correlations suppress noise-controlled switching and thus weaken the coherent response at the driving frequency.

We further examine $S$ over the range $\tau\in[0,1]$, as depicted in Fig.~\ref{fig:fig2}(b), where the right colorbar encodes the value of $S$. 
The same tendency persists across the full parameter plane: as $\tau$ increases, $S_{\mathrm{peak}}$ moves rightward and its height decreases. 
Overall, the largest amplification is attained in the white-noise limit ($\tau=0$), at $D_{\mathrm{optimal}}(\tau=0)$. 
Therefore, compared to Gaussian white noise, the colored noise suppresses SR in this higher-order network (1), as manifested by both a reduced $S_{\mathrm{peak}}$ and an increased $D_{\mathrm{optimal}}$ required to maximize $S$. 
Notably, this suppression is consistent with the well-known single-oscillator case  ($N=1$)~\cite{hanggi1993can}.

\subsection{Comparison of pairwise and higher-order networks}

\begin{figure}[!htb]
	\centering
	\begin{minipage}{0.49\linewidth}
		\centering
		\includegraphics[width=0.9\linewidth]{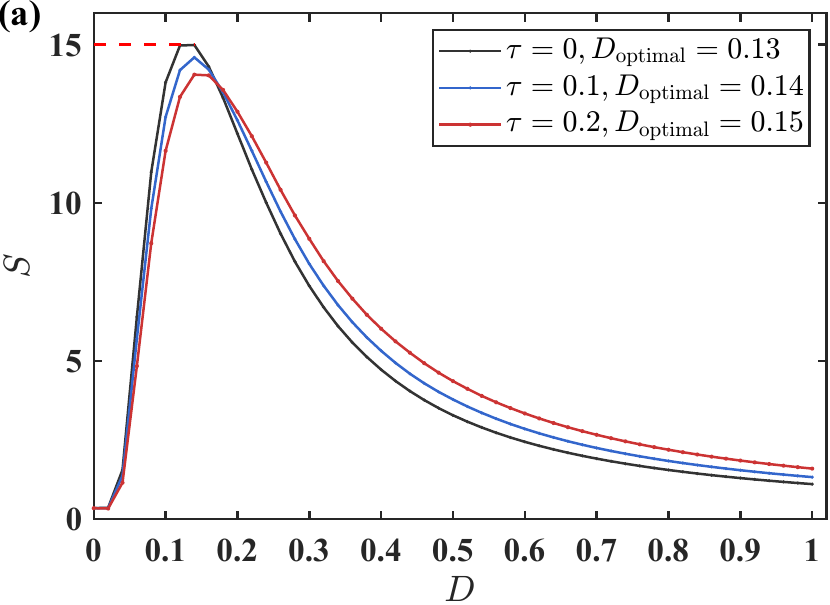}
	\end{minipage}
	\begin{minipage}{0.49\linewidth}
		\centering
		\includegraphics[width=0.9\linewidth]{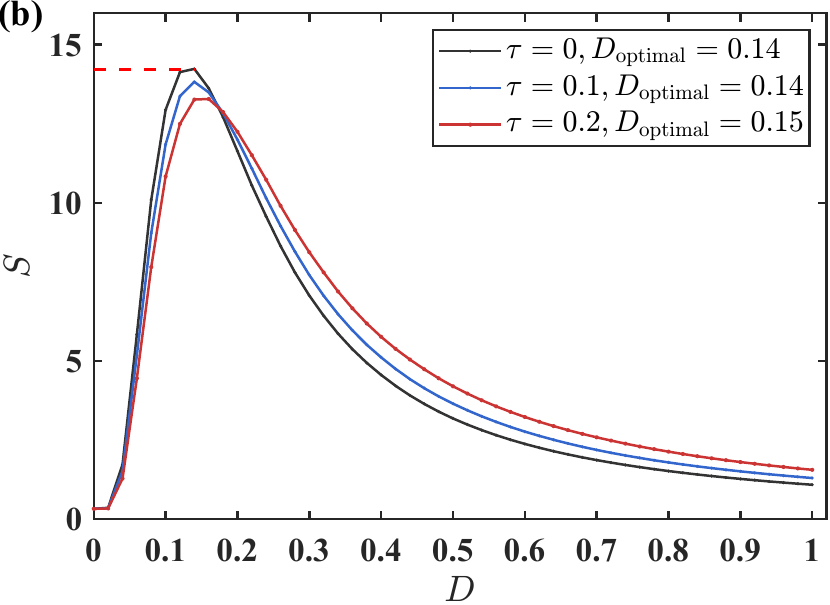}
	\end{minipage}

	\begin{minipage}{0.49\linewidth}
		\centering
		\includegraphics[width=0.9\linewidth]{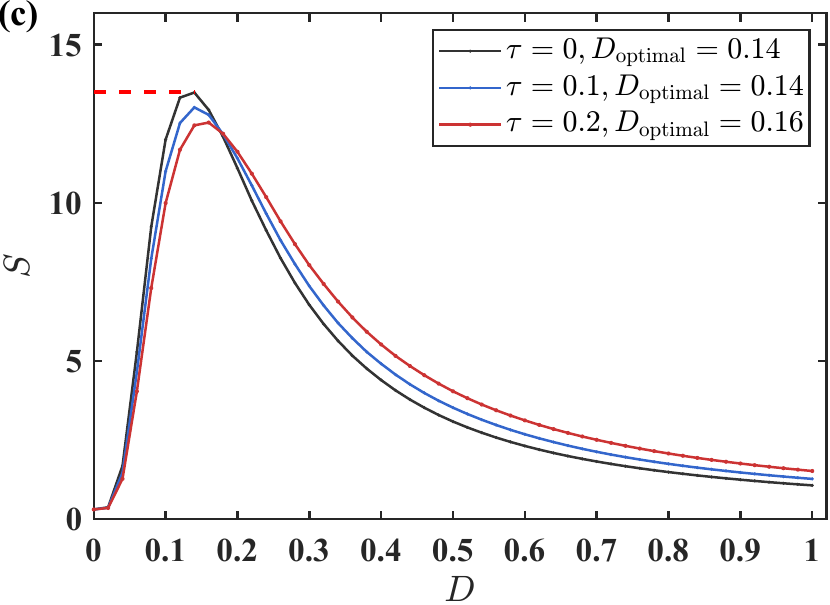}
	\end{minipage}
	\begin{minipage}{0.49\linewidth}
		\centering
		\includegraphics[width=0.9\linewidth]{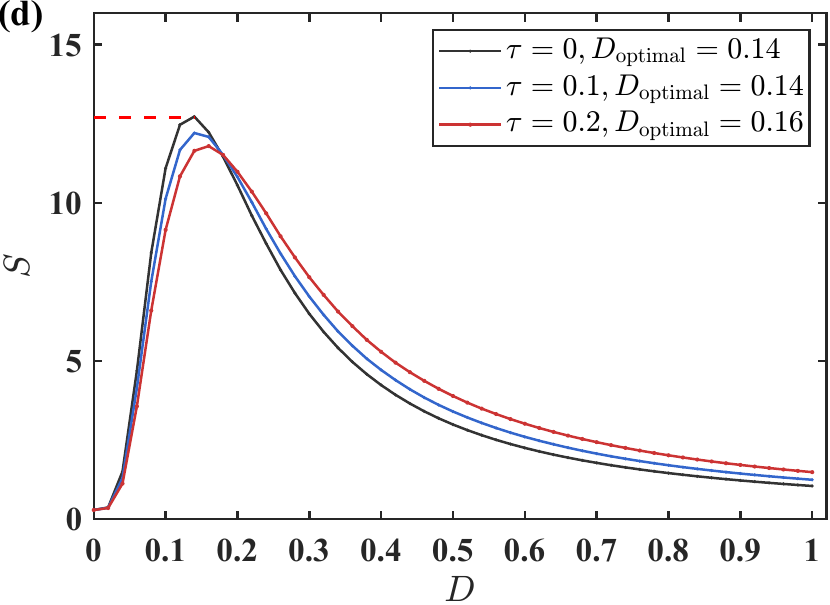}
	\end{minipage}

    \caption{Spectral amplification factor $S$ versus noise intensity $D$ for different weight $\alpha$ of higher-order interactions in the network (1). 
    (a) Purely pairwise network ($\alpha=0$); (b) $\alpha=0.25$; (c) $\alpha=0.5$; (d) $\alpha=0.75$. 
    For the white-noise case with the largest peak (best resonance) in each panel, the corresponding peak values indicated by red dashed lines are $S_{\mathrm{peak}}=15$, $14.2$, $13.5$, and $12.7$, respectively.}
    \label{fig:fig3}
\end{figure}

Figure~\ref{fig:fig3} compares SR across networks with different weights of higher-order (triadic) interactions controlled by $\alpha$. 
When $\alpha=0$, the system reduces to a purely pairwise-coupled network. The corresponding results are shown in Fig.~\ref{fig:fig3}(a). 
We observe a pronounced SR for all correlation times $\tau$ considered: $S(D)$ exhibits a clear unimodal dependence on $D$, with a well-defined $D_{\mathrm{optimal}}(\tau)$. 
Moreover, for each $\tau$, colored noise consistently suppresses SR, reflected by a lower peak and a rightward shift of the optimum.

For $\alpha=0.25$, Eq.~(\ref{eq:eq1}) incorporates both pairwise and triadic couplings. 
The results in Fig.~\ref{fig:fig3}(b) show that the same colored noise-induced suppression persists as $\tau$ varies. 
Figures~\ref{fig:fig3}(c) and \ref{fig:fig3}(d) present additional mixed-interaction cases ($\alpha=0.5$ and $\alpha=0.75$), and the qualitative behaviors remain unchanged: SR is retained, while increasing $\tau$ systematically weakens it.

In addition to the systematic suppression induced by increasing $\tau$, Fig.~\ref{fig:fig3} reveals a distinct effect of higher-order coupling itself.
As $\alpha$ increases, $S_{\mathrm{peak}}$ decreases monotonically, while $D_{\mathrm{optimal}}$ shifts slightly toward larger values at fixed $\tau$. 
In particular, the reported peak amplification decreases from $S_{\mathrm{peak}}=15$ at $\alpha=0$ to $S_{\mathrm{peak}}=12.7$ at $\alpha=0.75$, indicating a systematic reduction of coherence at the driving frequency as the triadic coupling becomes stronger. 
Concurrently, the rightward shift of $D_{\mathrm{optimal}}$ implies that stronger fluctuations are required to induce sufficiently frequent inter-well transitions in the presence of higher-order interactions.

Taken together, these observations suggest that higher-order interactions exert an overall inhibitory effect on SR in the present setting: they reduce the $S_{\mathrm{peak}}$ and shift the $D_{\mathrm{optimal}}$ to higher values. 
Crucially, the suppressive effect of colored noise remains qualitatively the same across all $\alpha$: for all $\alpha$, increasing the correlation time $\tau$ decreases $S_{\mathrm{peak}}$ and shifts $D_{\mathrm{optimal}}$ to larger values, consistent with the suppression effect observed in Fig.~\ref{fig:fig2}.

\subsection{The effects of coupling strength}
\begin{figure}[!htb]
	\centering
    \includegraphics[width=0.3\textwidth]{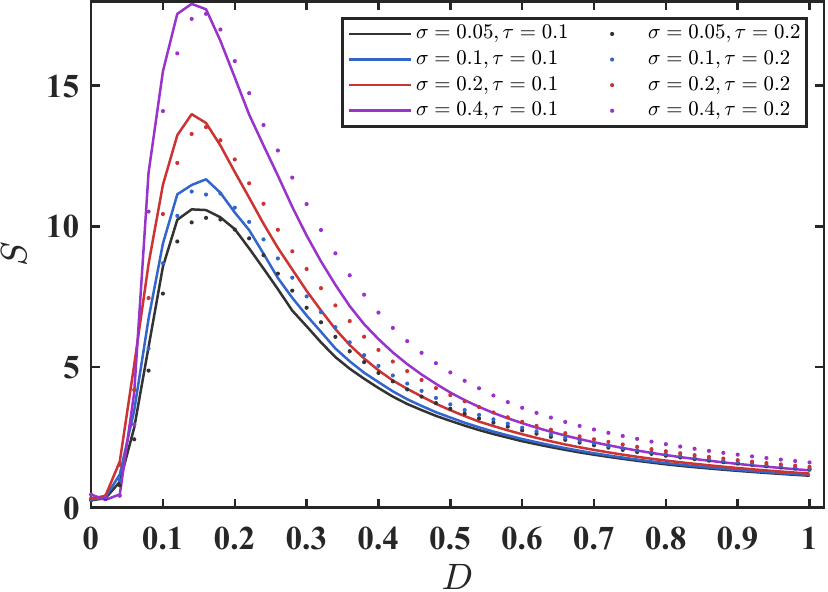}
	\caption{Spectral amplification factor $S$ versus noise intensity $D$ for different coupling strengths $\sigma$ on the higher-order network ($\alpha=1$). Four values of the coupling strength are shown: $\sigma=0.05,\,0.1,\,0.2,\,0.4$. The solid and dotted lines indicate the results of $\tau=0.1$ and $\tau=0.2$, respectively.}
    \label{fig:coupling}
\end{figure}
In coupled systems, it is generally expected that in the limit of strong coupling, one typically gets SR to be suppressed as the oscillators approach near-complete synchronization and the network effectively behaves as a low-dimensional collective variable. In this regime, the oscillators behave as a single collective variable, and SR tends to be suppressed as the system becomes dominated by the colored noise suppression effects characteristic of single-oscillator settings \cite{hanggi1993can}. However, the response of the system (\ref{eq:eq1}) outside the strong-coupling limit, particularly in higher-order networks, needs further investigation.

We examine the influence of the weak coupling strength $\sigma$ on SR, as shown in Fig.~\ref{fig:coupling}. 
These solid curves correspond to $\tau=0.1$ and dotted curves correspond to $\tau=0.2$, while different colors denote different values of $\sigma$. 
For each fixed $\tau$, the spectral amplification factor $S(D)$ exhibits a single resonance peak, from which we extract $S_{\mathrm{peak}}$ and $D_{\mathrm{optimal}}$. Our results show that for both values of $\tau$, increasing $\sigma$ enhances SR, leading to a higher $S_{\mathrm{peak}}$ and a more pronounced resonance profile. 
This enhancement is consistent with the standard array-enhanced SR mechanism \cite{lindner1996scaling}: a stronger coupling improves coherence among units and promotes more coordinated noise-induced barrier crossings, thereby strengthening the phase-locked response to the periodic forcing. At fixed $\sigma$, a larger $\tau$ results in a lower $S_{\mathrm{peak}}$ and a larger $D_{\mathrm{optimal}}$.  Therefore, our findings demonstrate that increasing $\sigma$ in the moderate range effectively promotes SR, while increasing the $\tau$ of colored noise suppresses it.
\subsection{The effects of power-limited colored noise}
\begin{figure}[!h]
	\centering
	\includegraphics[width=0.45\columnwidth]{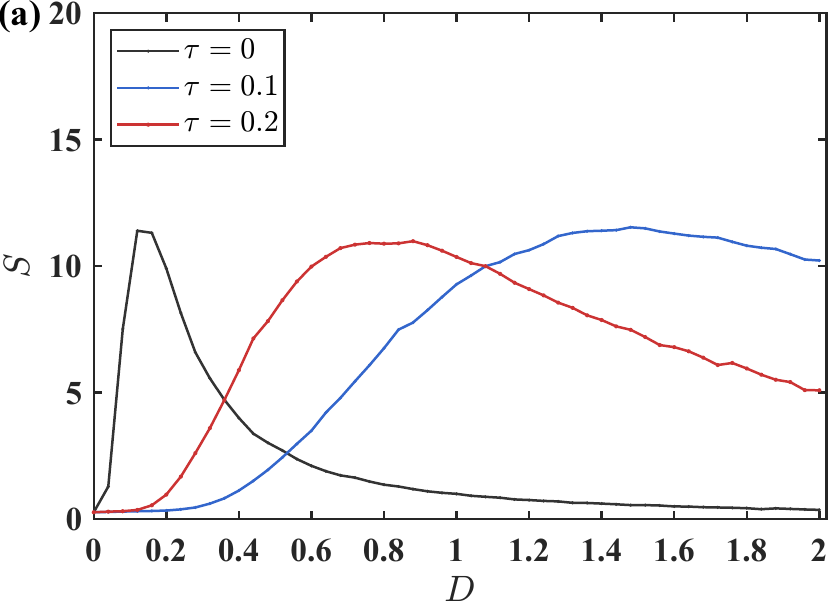}\quad
    \includegraphics[width=0.5\columnwidth]{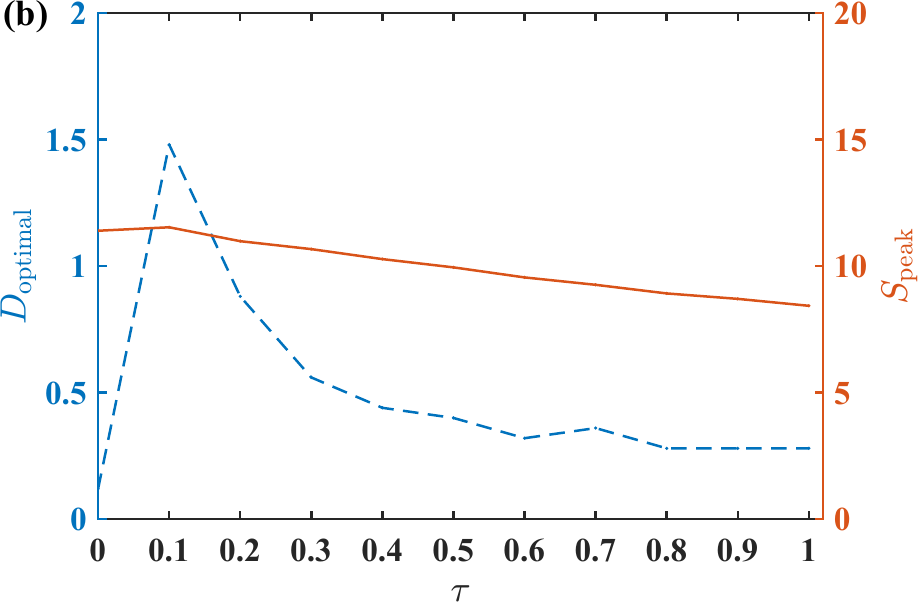}
	\caption{Effect of the correlation time $\tau$ on the spectral amplification factor in higher-order networks driven by power-limited colored noise. (a) $S$ versus $D^{\mathrm{pl}}$ for $\tau=0,\,0.1,\,0.2$; (b) $D^{\mathrm{pl}}_{\mathrm{optimal}}$ and $S_{\mathrm{peak}}$  versus $\tau$. Other parameters are the same as in Fig.~\ref{fig:fig2}.}
    \label{fig:fig7}
\end{figure}

We now investigate SR under the power-limited colored noise defined in Sec.~\ref{sec:sec2}. This analysis provides a direct comparison between the two OU parameterizations in the presence of higher-order coupling. The behavior of spectral amplification factor $S$ under power-limited colored noise is given in Fig.~\ref{fig:fig7}.

Figure~\ref{fig:fig7}(a) shows $S$ for different $\tau$. 
For each $\tau$, $S(D^{\mathrm{pl}})$ displays a clear unimodal dependence on $D^{\mathrm{pl}}$, confirming that SR persists under power-limited colored noise. 
When $\tau=0$, the optimal noise level $D^{\mathrm{pl}}_{\mathrm{optimal}}$ is relatively small. 
As $\tau$ increases, the peak height $S_{\mathrm{peak}}$ remains approximately unchanged over the explored range, whereas the optimal noise level $D^{\mathrm{pl}}_{\mathrm{optimal}}$ varies non-monotonically with $\tau$ (Fig.~\ref{fig:fig7}(b)). 
Therefore, power-limited colored noise primarily reshapes the resonance condition by shifting the optimal noise level, while leaving the peak amplification nearly intact in our parameter range.
This contrasts with the classical colored-noise case, where increasing $\tau$ simultaneously lowers $S_{\mathrm{peak}}$ and shifts $D^{\mathrm{pl}}_{\mathrm{optimal}}$ to larger noise intensities (Fig.~\ref{fig:fig2}).

Taken together, both OU colored-noise parameterizations reduce the efficiency of noise-controlled synchronization in the sense that increasing $\tau$ drives the system away from the optimal matching between the external forcing timescale and the noise-induced switching timescale.
The dominant manifestation, however, differs between those cases: under the classical case, increasing $\tau$ mainly lowers the resonance peak and shifts $D_{\mathrm{optimal}}$ to larger values, whereas under the power-limited case, $S_{\mathrm{peak}}$ remains approximately unchanged, while the optimal noise level is reshaped, yielding a pronounced non-monotonic dependence $D^{\mathrm{pl}}_{\mathrm{optimal}}(\tau)$.
\section{Understanding SR from a synchronization perspective}
\label{sec:sec4}
The results above raise a unifying question: what dynamical mechanism explains how the noise correlation time $\tau$ and the interaction structure jointly shape SR?
To address this, we reinterpret SR from the viewpoint of spatiotemporal synchronization, linking the resonance peak to an optimal compromise between temporal matching (noise-controlled switching versus periodic forcing) and spatial coherence (collective alignment across the networks).

Array-enhanced SR \cite{lindner1995array, lindner1996scaling} in coupled bistable oscillators is closely tied to spatiotemporal synchronization: the resonance peak typically occurs when temporal and spatial synchronization are simultaneously optimized. 
Temporal synchronization refers to phase locking between the external periodic forcing and noise-controlled inter-well switching, i.e., an optimal matching between the forcing period and the characteristic inter-well switching (waiting) time. 
Spatial synchronization quantifies the degree of collective coherence across oscillators, which is controlled by the coupling strength and the interaction structure, in particular including higher-order couplings.

In our setting, the mean-field variable $X(t)$ provides a macroscopic proxy for spatial coherence. 
When spatial synchronization is weak, oscillators populate the two wells incoherently, such that their contributions partially cancel and $X(t)$ stays close to zero, even though individual units may switch. 
Conversely, when spatial synchronization is strong, a large fraction of oscillators switch coherently, producing large-amplitude oscillations in $X(t)$.

In the remainder of this section, we analyze how higher-order coupling and colored-noise temporal correlations modulate temporal matching and spatial coherence, thereby shaping the peak $S_{\mathrm{peak}}$ and the optimal noise intensity $D_{\mathrm{optimal}}$ for the classical colored noise.

\subsection{Temporal matching quantified by mean-field switching}

SR can be understood as temporal synchronization: resonance emerges when the characteristic switching time matches the forcing timescale \cite{neiman1998stochastic}. 
For the forcing frequency $\Omega$, the standard matching condition is
$\pi/{\mathrm{MSWT}}\approx \Omega$,
i.e., the mean switching waiting time (MSWT) is close to half of the forcing period.

To quantify switching at the collective level, we compute the switching statistics of the mean-field variable $X(t)$. 
We record the times at which $X(t)$ crosses the separatrix $X=0$, and denote by $t^{1}_k$ and $t^{0}_k$ the $k$-th up-crossing ($X<0\to X>0$) and down-crossing ($X>0\to X<0$), respectively. 
Assuming that the first crossing is an up-crossing, the switching waiting time is defined as $\mathrm{SWT}_k=t^{0}_k-t^{1}_k,\quad t^{1}_k<t^{0}_k$. To avoid minor disturbances induced by noise near the threshold, we discard events with $\mathrm{SWT}_k<\pi/(10\Omega)$. 
The MSWT is then $\mathrm{MSWT}=\frac{1}{n}\sum_{k=1}^{n}\mathrm{SWT}_k$, and the corresponding mean switching frequency is
$\overline{\omega}=\pi/{\overline{\mathrm{MSWT}}}$, where $\overline{\mathrm{MSWT}}$ denotes the average MSWT over 50 independent realizations.

\begin{figure}[!htb]
	\centering
	\includegraphics[width=0.45\columnwidth]{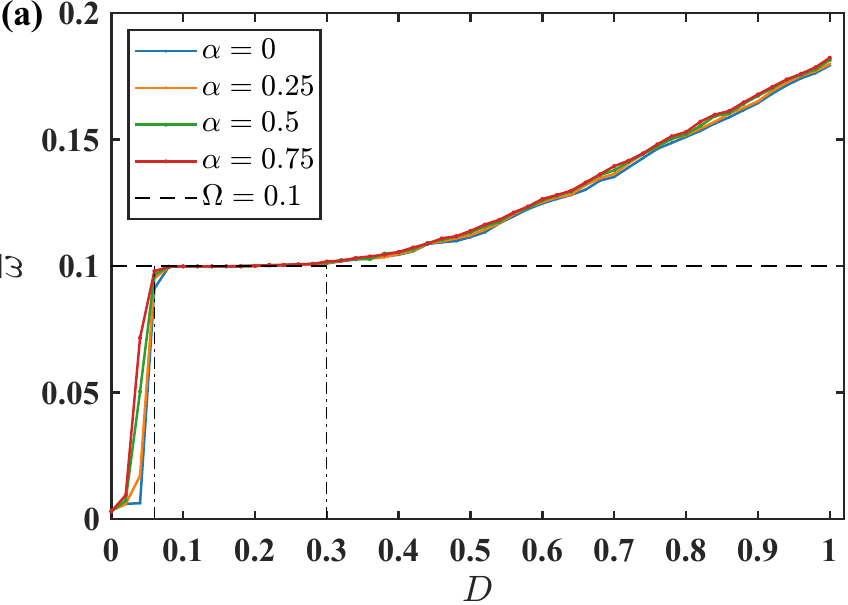}\quad
    \includegraphics[width=0.45\columnwidth]{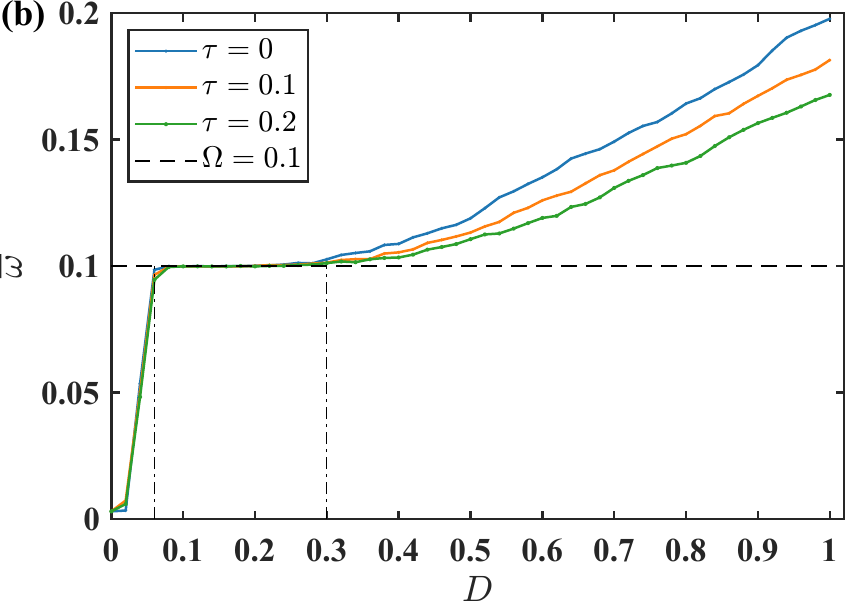}
	\caption{Mean switching frequency $\overline{\omega}$ versus noise intensity $D$. (a) With the correlation time fixed at $\tau=0.1$ for higher-order interaction weights $\alpha=0,\,0.25,\,0.5,\,0.75$; (b) With the higher-order interaction weight fixed at $\alpha=0.5$ for colored-noise correlation times $\tau=0,\,0.1,\,0.2$.  Other parameters are identical to those in Fig.~\ref{fig:fig2}.}
    \label{fig:resonance band}
\end{figure}

We then compare $\overline{\omega}(D)$ with the forcing frequency $\Omega$ to identify the noise range where temporal matching holds and to relate it to the resonance peak in $S(D)$. 
Figure~\ref{fig:resonance band} shows that $\overline{\omega}(D)$ exhibits three regimes: 
i) For subthreshold noise ($D<D_1\approx0.06$), $\overline{\omega}<\Omega$, indicates rare inter-well transitions and thus weak entrainment. 
ii) For intermediate noise ($D_1<D<D_2$ with $D_2\approx0.3$), $\overline{\omega}(D)$ forms a plateau close to $\Omega$, i.e., the matching condition is approximately satisfied and switching becomes phase locked to the external periodic forcing; the SR peak in Fig.~\ref{fig:fig2} lies within this matched-noise interval. 
iii) For strong noise ($D>D_2$), the switching becomes too rapid and irregular, and $\overline{\omega}$ grows beyond $\Omega$, reflecting a noise-dominated regime in which coherent entrainment is lost. 
These three regimes mirror the classical behavior of a single overdamped bistable oscillator, indicating that higher-order coupling does not qualitatively change the temporal-matching mechanism.

Figure~\ref{fig:resonance band}(a) further exhibits that varying the higher-order coupling weight $\alpha$ has only a weak influence on $\overline{\omega}(D)$, which is consistent with the observation that $\alpha$ mainly modulates SR through spatial coherence rather than through the switching timescale itself. By contrast, Fig.~\ref{fig:resonance band}(b) indicates that the effect of $\tau$ becomes visible primarily at strong noise ($D\gtrsim0.3$): larger $D$ slows down the growth of $\overline{\omega}$, reflecting the suppression of fast stochastic switching by temporal correlations. However, this temporal effect alone does not explain the reduction of $S_{\mathrm{peak}}$ and the shift of $D_{\mathrm{optimal}}$ observed earlier, motivating an explicit analysis of spatial synchronization in the next subsection.

\subsection{Relationship between SR and network synchronization}

Temporal matching alone does not fully determine the mean-field SR of the networks: the level of synchronization among oscillators (spatial coherence) is equally crucial. 
Indeed, array-enhanced SR is maximized only when temporal matching and spatial coherence are simultaneously optimized \cite{lindner1995array}. 
If spatial synchronization is weak, even an appropriate noise level may not yield a large mean-field response because the population is 
not simultaneously in the same-side potential well. 
Conversely, strong spatial synchronization without temporal matching leads to coherent but off-resonant switching, which also reduces the spectral amplification factor at the driving frequency.

This motivates a complementary viewpoint in which SR is interpreted through network synchronization: the resonance peak in $S(D)$ should be accompanied by an extrema of a suitable spatial-synchronization measure. 
Within this framework, higher-order coupling and colored-noise correlations affect SR primarily by reshaping the $D$-dependence of spatial coherence, thereby shifting the location and height of the resonance peak. 
Specifically, we examine how the higher-order coupling weight $\alpha$ and the noise parameters $(D,\tau)$ modulate the network’s spatial synchronization, and we relate the resulting extrema to the resonance peak and the shift of $D_{\mathrm{optimal}}$.

We use the standard deviation $R$ of node states to quantify spatial synchronization, defined as
\begin{equation}
	R =E\!\left(\left<\mathrm{std}(x_i(t))\right>\right),\quad 
\mathrm{std}(x_i(t))=\sqrt{\frac{1}{N} \sum_{i=1}^{N}\bigl(x_i(t)-X(t)\bigr)^2},
\end{equation}
where $\langle\cdot\rangle$ denotes time averaging and $E(\cdot)$ is estimated by averaging over realizations. 
Smaller $R$ corresponds to stronger synchronization, and $R=0$ indicates perfect synchronization.

\begin{figure}[!htb]
	\centering
	\includegraphics[width=0.4\textwidth]{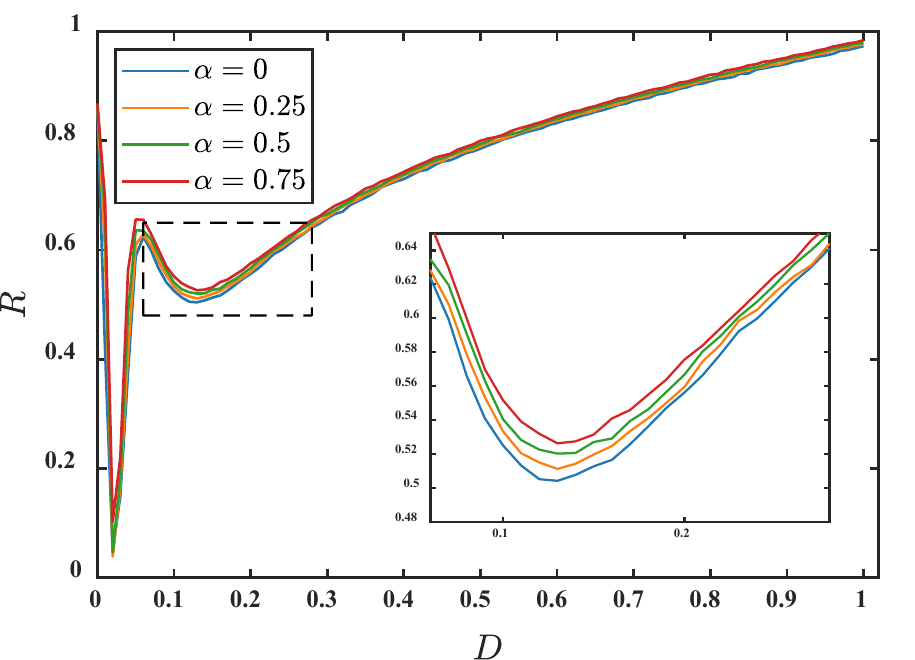}
	\caption{The curves of spatiotemporal synchronization level $R$ versus noise intensity $D$. With the correlation time fixed at $\tau=0.1$ for higher-order interaction weights $\alpha=0, 0.25, 0.5, 0.75$. Other parameters are identical to those in Fig. 2.}
    \label{fig:synchronization}
\end{figure}

Figures~\ref{fig:synchronization} and 8(a) plot $R$ as a function of the noise intensity $D$. 
The effects of the higher-order interactions weight $\alpha$ and the correlation time $\tau$ on $R(D)$ relate with the resonant curves $S(D)$: the parameter changes such that lower SR peaks are accompanied by weaker spatial coherence. 
Overall, $R(D)$ exhibits a clear Four-stage variation in network synchronization level under different $\alpha$ and
$\tau$.

Using Fig.~\ref{fig:synchronization} as an illustrative example, we elaborate on the variation of network synchronization level across four stages in detail. To interpret this non-monotonic behavior, we first consider the baseline case of $D=0$. 
In the absence of noise, each overdamped oscillator rapidly relaxes into the potential well determined by its initial condition. 
With initial values drawn uniformly from $[-1,1]$, the oscillators partition between the two stable wells (around the stable equilibria $x_p=\pm1$), yielding an approximately balanced two-well population and hence weak spatial coherence.

When a small amount of noise is introduced, rare barrier-crossing events break this fragile balance by creating a slight majority in one of the two wells (Stage~I, $D\in(0,0.02]$). 
Because switching remains infrequent at such weak noise, this imbalance persists long enough for coupling to amplify it, ultimately pulling most oscillators into the same well. 
As a result, the network exhibits strong spatial synchronization and $R$ reaches a first minimum.

As $D$ increases further, noise-induced transitions become more frequent and progressively undermine this single-well alignment, leading to an increase of $R$ (Stage~II, $D\in(0.02,0.06]$). The coupling can now no longer suppress noise-induced transitions.
Upon entering the intermediate-noise regime, where the mean switching time becomes comparable to half of the forcing period, an increasing fraction of oscillators begins to switch in a phase-locked manner (Stage~III, $D\in(0.06,0.15]$ ). 
Coupling then facilitates the recruitment of the remaining oscillators, restoring coherence during collective switching and driving $R$ toward a second minimum. 
For strong noise (Stage~IV, $D>0.15$), switching becomes rapid and irregular, and spatial coherence is gradually destroyed, producing an overall increase of $R$.

Comparing Fig.~\ref{fig:synchronization} with Fig.~\ref{fig:fig2}, we find that the SR peak occurs near the second minimum of $R(D)$. 
That is, the noise level that maximizes the spectral amplification factor typically coincides with the noise level at which the network exhibits the strongest spatial coherence during collective switching. 
This provides a direct synchronization-based explanation why SR is strongest only when temporal matching and spatial coherence are achieved simultaneously.

\begin{figure}[!htb]
	\centering
	\includegraphics[width=0.5\textwidth]{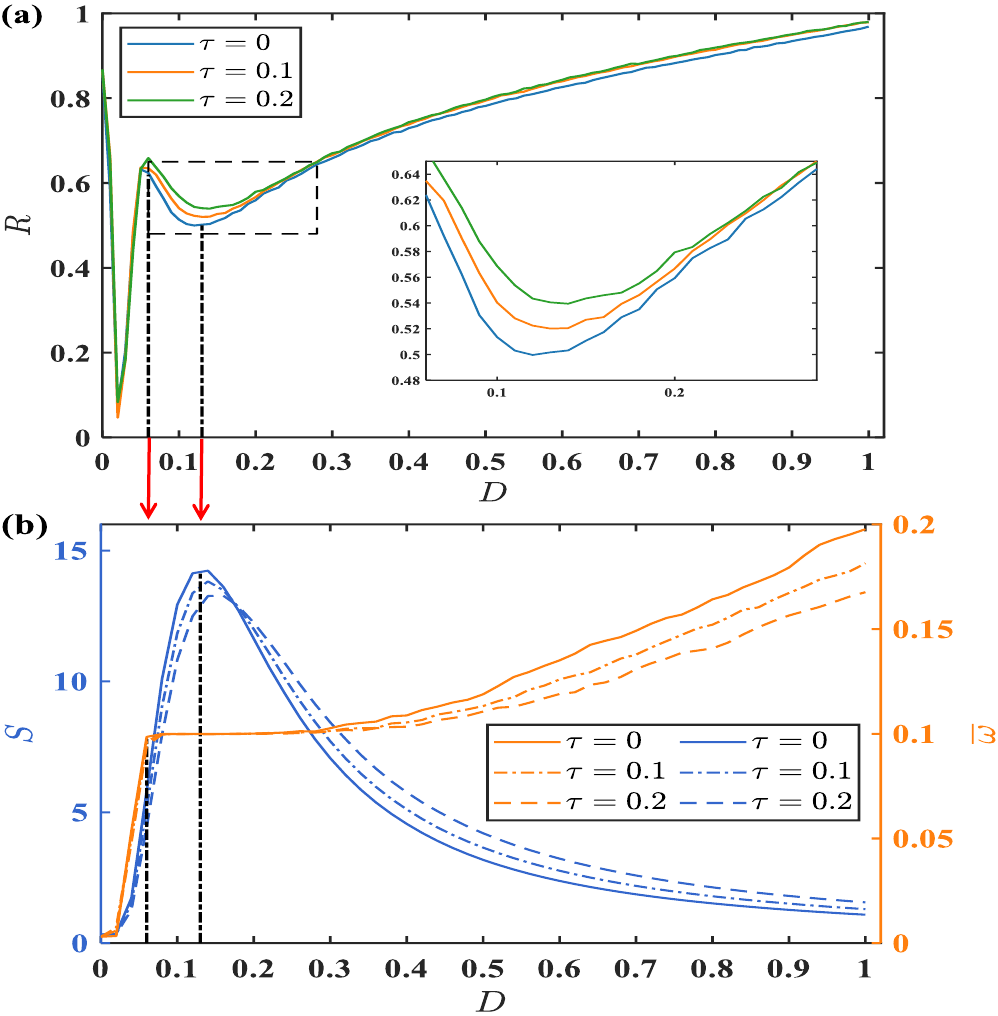}
	\caption{The curves of spatiotemporal synchronization level $R$ versus noise intensity $D$. (a) With the higher-order interaction weight fixed at $
\alpha=0.5$ for colored noise correlation time $\tau=0, 0.1, 0.2$; (b) Dual-Y plot showing the spectral amplification factors $S$ (left axis, blue, copy from Fig.~\ref{fig:fig3}(b)) and the mean switching frequency $\overline{\omega}$ (right axis, orange, copy from Fig.~\ref{fig:resonance band}(b)). The corresponding relations of the extrema in Fig.~\ref{fig:relations}(a) and (b) are marked by two identical noise intensity (two black vertical dash-dotted lines).}
    \label{fig:relations}
\end{figure}

This correspondence also clarifies the roles of higher-order coupling and colored noise correlations. 
Increasing $\alpha$ raises the minimum value of $R$ attained near resonance and shifts its location slightly toward larger $D$, implying weaker spatial coherence and hence a reduced $S_{\mathrm{peak}}$ together with a rightward shift of $D_{\mathrm{optimal}}$. 
Similarly, Fig.~\ref{fig:relations} shows the effects of $\tau$ on spatiotemporal synchronization $R$, which are linked to the spectral amplification factors $S$ and the mean switching frequency $\overline{\omega}$. Increasing $\tau$ weakens the peak of spatial coherence (larger R) around the resonant switching regime and displaces the synchronization minimum to larger $D$, consistently accounting for the reduced $S_{\mathrm{peak}}$ and the increased $D_{\mathrm{optimal}}$ observed under colored noise.

Overall, the synchronization measure $R(D)$ complements the temporal-matching analysis based on $\overline{\omega}(D)$. 
While $\overline{\omega}(D)\approx\Omega$ identifies the noise range, where phase locking is possible, the minimum of $R(D)$ pinpoints where spatial coherence is maximized, thereby locating $D_{\mathrm{optimal}}$ more precisely and explaining the variations of $S_{\mathrm{peak}}$ across $\alpha$ and $\tau$.

Some remarks on noise-intensity-modulated network synchronization curves are given: 
(1). Influence of initial conditions. When all oscillators are initialized in the same potential well, network coupling will maintain their stability in that well, effectively suppressing inter-well transitions. In this scenario, the Stage I, the synchronization enhancement is replaced by weakening, as $R(0)=0$. While the first extrema (at $D\approx 0.02$) vanishes under these conditions, the second extrema (at $D\approx 0.15$) remains unchanged. 
(2). In Fig.~\ref{fig:relations}, the extrema ($D\approx0.06$) of the $R(D)$ curve corresponds perfectly with temporal synchronization's trigger, marking where the globally coupled overdamped oscillator network begins amplifying input signals. 
(3). The extrema points of network synchronization precisely locate the optimal noise intensity for SR, whereas the mean switching frequency only identifies a noise interval (resonance band) rather than the exact optimum.

\section{Conclusions}

We study SR in globally coupled overdamped bistable oscillators under the combined influence of higher-order interactions and temporally correlated Gaussian noise. 
Our numerical results reveal that increasing either the noise correlation time $\tau$ or the higher-order coupling weight $\alpha$ suppresses SR, manifested by a reduced resonance peak and a rightward shift of the optimal noise intensity. 
Importantly, introducing higher-order coupling does not change the qualitative influence of colored noise: classical colored noise remains suppressive, and power-limited colored noise retains its characteristic non-monotonic dependence on $\tau$.

A synchronization-based interpretation clarifies the underlying mechanism of SR. 
The correlation time $\tau$ primarily acts on temporal matching by reshaping inter-well hopping statistics, whereas the higher-order interactions weight $\alpha$ mainly acts on spatial coherence by regulating how switching activity spreads across oscillators. 
Accordingly, the SR peak of $S(D)$ occurs near an extremum of the synchronization measure $R(D)$, indicating that maximal amplification requires temporal matching together with strong spatial coherence. 
Within the weak-coupling regime considered here, higher-order coupling therefore modulates the collective coherence of switching but does not overturn the fundamental role of colored noise in controlling hopping dynamics.

An interesting direction for future work is to identify counterexamples in overdamped symmetric bistable networks where higher-order coupling reverses, rather than preserves, the classical colored-noise suppression of SR. 
Higher-order interactions substantially broaden the design space, offering diverse higher-order coupling functions and higher-order network topologies that may enable such behaviors. 
More broadly, our results help to  clarify collective dynamics jointly shaped by colored noise and higher-order interactions—two ingredients frequently invoked in models of neuronal population activity in the brain.

\section*{Author contributions}

Z.B. and D.Z. contributed equally to this paper.

Zhongwen Bi: Conceptualization (equal); Formal analysis (equal); Investigation (equal); Methodology
(equal); Visualization (equal); Writing – original draft (equal); Writing – review \& editing (equal). Dan Zhao: Conceptualization (equal); Formal analysis (equal); Investigation (equal); Writing –
original draft (equal); Writing – review \& editing (equal). Qi Liu: Investigation (equal); Writing – review \&
editing (equal). J\"urgen Kurths: Conceptualization (equal); Supervision (equal); Writing – review \& editing (equal). Yong Xu: Conceptualization (equal); Supervision (equal); Writing – review \& editing (equal).

\section*{Declaration of competing interest}
The authors declare that they have no known competing financial interests or personal relationships that could have appeared to influence the work reported in this paper.

\section*{Data availability}
The data that supports the findings of this study are available within the article.

\section*{Acknowledgements}
\begin{sloppypar}
This work was partly supported by the NSF of China (Grant No.~12120101002). D. Zhao thanks
the Sino-German (CSC-DAAD) Postdoc Scholarship Program. Q. Liu gratefully acknowledges the support of the China Scholarship Council (Grant No.~202006290313).
\end{sloppypar}

%

\end{document}